# Sound-to-Vibration Transformation for *Sensorless* Motor Health Monitoring

Ozer Can Devecioglu, Serkan Kiranyaz, Amer Elhmes, Sadok Sassi, Turker Ince, Onur Avci, Mohammad Hesam Soleimani-Babakamali, Ertugrul Taciroglu, and Moncef Gabbouj, *Fellow, IEEE*

*Abstract*— Automatic sensor-based detection of motor failures such as bearing faults is crucial for predictive maintenance in various industries. Up to 51% of motor failures are attributed to bearing faults alone. Such failures can lead to unexpected downtime, increased maintenance costs, and even catastrophic accidents. Numerous methodologies have been developed over the years to detect bearing faults. Despite the appearance of numerous different approaches for diagnosing faults in motors have been proposed, vibration-based methods have become the de facto standard and the most commonly used techniques. However, acquiring reliable vibration signals, especially from rotating machinery, can sometimes be infeasibly difficult due to challenging installation and operational conditions (e.g., variations on accelerometer locations on the motor body), which will not only alter the signal patterns significantly but may also induce severe artifacts. Moreover, sensors are costly and require periodic maintenance to sustain a reliable signal acquisition. To address these drawbacks and void the need for vibration sensors, in this study, we propose a novel sound-to-vibration transformation method that can synthesize realistic vibration signals directly from the sound measurements regardless of the working conditions, fault type, and fault severity. As a result, using this transformation, the data acquired by a simple sound recorder, e.g., a mobile phone, can be transformed into the vibration signal, which can then be used for fault detection by a pre-trained model. The proposed method is extensively evaluated over the benchmark Qatar University Dual-Machine Bearing Fault Benchmark dataset (QU-DMBF), which encapsulates sound and vibration data from two different machines operating under various conditions. Experimental results show that this novel approach can synthesize such realistic vibration signals that can directly be used for reliable and highly accurate motor health monitoring. The benchmark dataset, our results, and the optimized PyTorch implementation of the proposed approach are now publicly available.

## I. INTRODUCTION

MOTOR fault detection is an essential component of a predictive maintenance pipeline in various industries, such as manufacturing, aerospace, and energy. Bearings are essential components of rotating machinery equipment which is commonly used in industry, and their failure can result in unexpected downtime, increased maintenance costs, and even catastrophic accidents. Numerous methods and techniques have been presented to recognize and address bearing problems throughout the years, which has led to substantial research into bearing fault identification. These methods can be grouped into three main categories: model-based methods [1]-[4], traditional signal-processing approaches, [5]-[11], and machine (ML) and deep learning (DL) methods, [12]-[24]. A multitude of bearing fault detection methods was proposed in the last decade, and the above-mentioned techniques all share the common trait of using vibration data to identify certain potential faults. Due to its capacity to track changes efficiently in the mechanical behavior of the bearing, vibration data is commonly used for fault detection and thus set a *de facto* standard in this domain. As bearings begin to degrade, their vibration signature changes. By examining the variations in the vibration signature, it is possible to identify the early stages of bearing failure and take corrective action before a catastrophic failure takes place. DL approaches have successfully been used to detect faults in rotating machines directly over raw vibration data [12]-[25]. However, certain problems may occur in real-time implementations of such fault detection methods. First, reliable vibration data acquisition, especially from a rotating machine at high speeds can be a challenging task. Electric motors can produce high levels of background noise including electrical interference, ambient vibrations, and sensor noise, which frequently alter the vibration signals. Due to this noise, vibration readings may not be accurate, resulting in false alarms or missed detections.

O. Devecioglu and M. Gabbouj are with the Department of Computing Sciences, Tampere University, Tampere, Finland (e-mail: ozer.devecioglu@tuni.fi, moncef.gabbouj@tuni.fi,).
S. Kiranyaz, is with the Electrical Engineering Department, Qatar University, Doha, Qatar (e-mail: mkiranyaz@qu.edu.qa). A. Elhmes, and S. Sassi are with the Civil Engineering Department, Qatar University, Doha, Qatar (e-mail: aa1702913@qu.edu.qa; sadok.sassi@qu.edu.qa)
T. Ince, is with the Electrical and Electronics Engineering Department, Izmir University of Economics, Izmir, Turkey (email:turker.ince@ieu.edu.tr).

O. Avci is with the Department of Civil, Construction and Environmental Engineering, Iowa State University, Ames, IA, USA (email: onur.avci@mail.wvu.edu )
MH. Soleimani-Babakamali and E. Taciroglu are with the Department of Civil and Environmental Engineering, University of California, Los Angeles, CA, USA. (email: soleimanisam92@ucla.edu , etacir@ucla.edu )

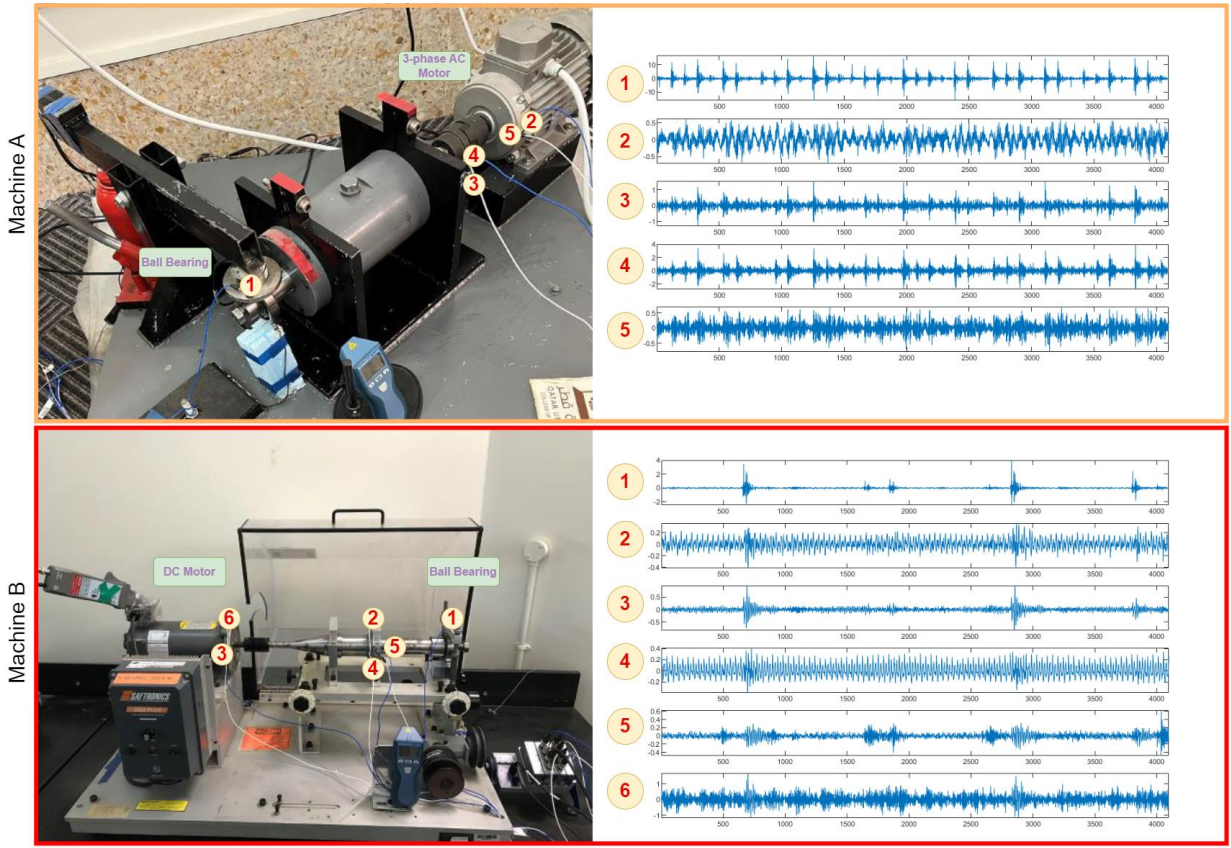

**Figure 1** Samples of *simultaneously* recorded vibration signals over different sensor locations on the QU-DMBF dataset.

Besides acquisition-induced problems, installing accelerometers with wires in certain locations of rotating machines may also pose certain difficulties and operational drawbacks. Furthermore, the vibration sensors, e.g., accelerometers, may not work well or even break under harsh conditions. On the other hand, wireless accelerometers, especially with a high-resolution data capability are quite costly and require periodic maintenance for a reliable acquisition. Another critical issue is the mounting location of vibration sensors where slight positional variations cause significant changes in vibration patterns. Figure 1 shows raw vibration signals from the two distinct machines of the QU-DMBF dataset [25]. For each machine, the vibration signals are acquired by distinct accelerometers within a close proximity. It is clear in Figure 1 that signals from different accelerometers are entirely different from one another. For instance, even though sensors 2 and 5 from Machine A and sensors 6 and 3 as well as 2, 5, and 4 from Machine B are placed close to each other, their vibration signals are entirely different. Therefore, a DL model trained over one of these sensors for fault detection will ultimately underperform or may even fail on the data acquired from another sensor placement.

To address all the aforementioned drawbacks, this study proposes sound-to-vibration transformation with the aim of *sensorless* fault detection. Consider the following practical use-case scenario: once the fault detector is trained over the vibration data acquired by an accelerometer at any location, the proposed transformer can also be trained over the vibration (from the same sensor) and sound data by the manufacturer of the motor. The proposed transformer thus learns to synthesize realistic vibration data directly from the acquired sound measurements. The two trained models, the fault detector and the transformer will then be shared with the motor's operator for continuous health monitoring. In this way, realistic vibration signals can be produced directly from the sound (recorded by e.g., a mobile phone) and then the pre-trained detector over the synthesized vibration data can be used for fault detection during the operational lifetime of the motor as illustrated in Figure 2. Therefore, machine operators will no longer need to purchase, install, and maintain accelerometer(s) with the *same* setup used by the manufacturer (e.g., the *same* sensor model installed at the *same* location during the training of the fault detector).

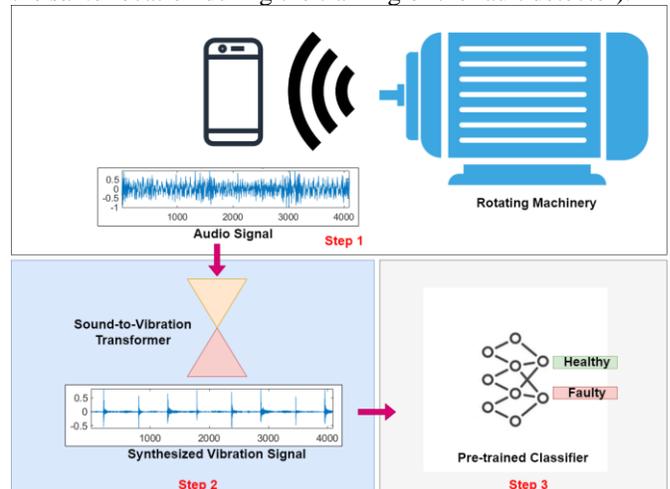

**Figure 2** A practical use-case illustration of the proposed motor fault detection.

As the typical use-case scenario shown in Figure 2, besides the cost and energy savings, the proposed approach will also yield a robust fault detection since it can synthesize a vibration signal that is similar to the one used during the training of the classifier by the manufacturer regardless of the aforementioned variations. This makes fault detection far more reliable and accurate by a classifier that was pre-trained over the actual vibration signal.

In this study, 1D Operational U-Net (Op-UNet) is used as the network model for the proposed sound-to-vibration transformer. Op-UNets have Self-Organized Operational Neural Network (Self-ONN) architecture [27]-[33] which are heterogeneous network models with generative neurons that can perform optimal non-linear operations for each kernel element. Self-ONNs have outperformed their CNN counterparts in many tasks [24], [25], [34]-[41] significantly even with a reduced network complexity and depth compared to CNNs. In this study, we aim to leverage this superiority to synthesize highly realistic vibration signals and thus achieve the same fault detection performance level by the manufacturer's pre-trained model over the original vibration signal.

The rest of the paper is organized as follows: a brief outline of 1D Self-ONNs and the proposed sound-to-vibration transformer with the Operational U-Nets are introduced in Section II. A detailed set of experimental results over the two distinct machines from the QU-DMBF dataset are presented in Section III. Finally, Section IV concludes the paper and suggests topics for future research.

## II. METHODOLOGY

### A. 1D Self-Organized Operational Neural Networks

In this section, we briefly review Self-ONNs with some of their key characteristics. Different from the convolution operator of CNNs, the nodal operator of each generative neuron of a Self-ONN can perform any nonlinear transformation which can be expressed based on Taylor approximation near origin:

$$\psi(x) = \sum_{n=0}^{\infty} \frac{\psi^{(n)}(0)}{n!} x^n \quad (1)$$

The $Q^{th}$ order truncated approximation, formally known as the Taylor polynomial, is represented by the following finite summation:

$$\psi(x)^{(Q)} = \sum_{n=0}^{Q} \frac{\psi^{(n)}(0)}{n!} x^n \quad (2)$$

The above formulation can approximate any arbitrary function $\psi(x)$ near 0. When the activation function bounds the neuron's input feature maps in the vicinity of 0 (e.g., *tanh*), the formulation in (2) can be exploited to form a composite nodal operator where the power coefficients, $\frac{\psi^{(n)}(0)}{n!}$, can be the parameters of the network learned during training.

It was shown in [35]-[37] that the 1D nodal operator of the $k^{th}$ generative neuron in the $l^{th}$ layer takes the following general form:

$$\widetilde{\psi_k^l}\left(w_{ik}^{l(Q)}(r), y_i^{l-1}(m+r)\right)$$
$$= \sum_{q=1}^{Q} w_{ik}^{l(Q)}(r,q) \left(y_i^{l-1}(m+r)\right)^q \quad (3)$$

Let $x_{ik}^l \in \mathbb{R}^M$ be the contribution of the $i^{th}$ neuron at the $(l-1)^{th}$ layer to the input map of the $l^{th}$ layer. Therefore, it can be expressed as,

$$\widetilde{x_{ik}^l}(m) = \sum_{r=0}^{K-1} \sum_{q=1}^{Q} w_{ik}^{l(Q)}(r,q) \left(y_i^{l-1}(m+r)\right)^q \quad (4)$$

where $y_i^{l-1} \in \mathbb{R}^M$ is the output map of the $i^{th}$ neuron at the $(l-1)^{th}$ layer, $w_{ik}^{l(Q)}$ is a learnable kernel of the network, which is a $K \times Q$ matrix, i.e., $w_{ik}^{l(Q)} \in \mathbb{R}^{K \times Q}$, formed as, $w_{ik}^{l(Q)}(r) = [w_{ik}^{l(Q)}(r,1), w_{ik}^{l(Q)}(r,2), ..., w_{ik}^{l(Q)}(Q)]$. By the commutativity of the summation operations in (4), one can alternatively write:

$$\widetilde{x_{ik}^l}(m) = \sum_{q=1}^{Q} \sum_{r=0}^{K-1} w_{ik}^{l(Q)}(r, q-1) y_i^{l-1}(m+r)^q \quad (5)$$

One can simplify this as follows:

$$\widetilde{x_{ik}^l} = \sum_{q=1}^{Q} Conv1D\left(w_{ik}^{l(Q)}, \left(y_i^{l-1}\right)^q\right) \quad (6)$$

Hence, the formulation can be accomplished by applying Q 1D convolution operations. Finally, the output of this neuron can be formulated as follows:

$$x_k^l = b_k^l + \sum_{i=0}^{N_{l-1}} x_{ik}^l \quad (7)$$

where $b_k^l$ is the bias associated with this neuron. The $0^{th}$ order term, $q = 0$, the DC bias, is ignored as its additive effect can be compensated by the learnable bias parameter of the neuron. With the $Q = 1$ setting, a *generative* neuron reduces back to a convolutional neuron.

The raw-vectorized formulations of the forward propagation and detailed formulations of the Back-Propagation (BP) training in the raw-vectorized form can be found in [27], [29], and [35].

### B. Sound to Vibration Transformation by 1D Operational UNet

The ultimate goal of this study is to synthesize highly realistic vibration signals of a motor from its sound so that the synthesized vibration data can directly be used for accurate fault detection by the original classifier pre-trained on the actual vibration data. As discussed earlier, this will void the need for any sensor installation, and a simple sound recorder will suffice for continuous motor health monitoring.

In this study, 1-second paired audio and vibration signals are used to train the Operational U-Net. Each segment is first linearly normalized as follows:

$$X_N(i) = \frac{2(X(i) - X_{\min})}{X_{max} - X_{min}} - 1 \quad (8)$$

where $X(i)$ is the original sample amplitude in the segment, $X_N(i)$ is the normalized segment, $X_{min}$ and $X_{max}$ are the minimum and maximum amplitudes within the segment, respectively. This will scale the segment linearly in the range of [-1 1] where $X_{min} \rightarrow -1$ and $X_{max} \rightarrow 1$.

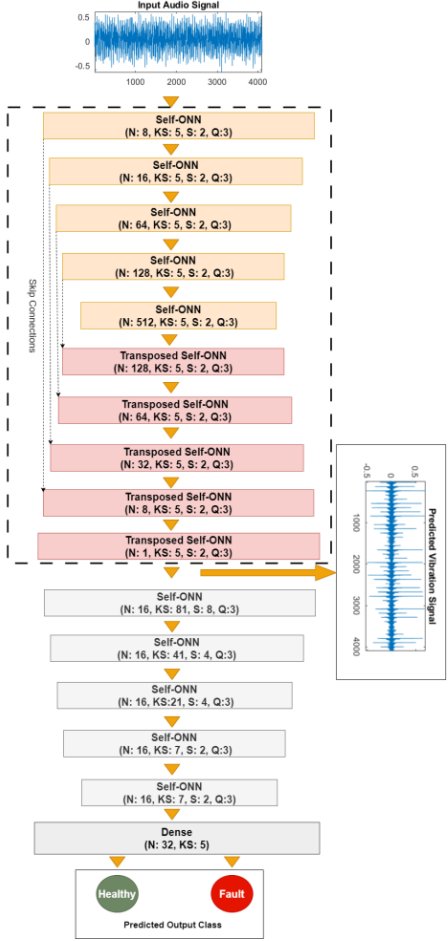

**Figure 3:** The network architecture of the Op-UNet model.

As illustrated in Figure 3, the proposed network has a total of 15 operational layers. The first 10 layers are organized into a 1D Operational U-Net (Op-UNet) model which consists of 5 operational layers in the encoder and 5 transposed operational layers in the decoder with skip connections. This is the transformer network, where the input and output of the network are the 1-second sound and the corresponding vibration segments, respectively. Figure 4 depicts the training process of the Op-UNet. To reinforce the learning by emphasizing the fault status of the given signal, in addition to this 10-layer U-Net, the transformer network is cascaded with a Self-ONN classifier with 5 operational layers and 2 dense layers. By cascading this classifier to the transformer, the training will use the advantage of discriminating *fault* segments from *normal* segments, which in turn will improve the regression (transformation) task. After training, the classifier can be ignored.

The objective function used for training consists of a combination of three distinct loss functions. To generate more realistic vibration signals, both temporal and spectral signal representations are taken into consideration by utilizing their corresponding loss functions. The objective is, therefore, to minimize the mean-absolute error (MAE) in both time and frequency domains.

The MAE loss function in the *time* domain, where, $X_N$ is the normalized input *sound* signal, $Synth(X_N)$ is the synthesized *vibration* signal and $Y$ is the corresponding actual vibration signal can be formulated as follows:

$$Loss_{Time} = \|(Y - Synth(X_N))\|_1 \quad (9)$$

For the spectral loss function, the N-point discrete STFT of the input and output signals is first performed as expressed in Eq. (10), where X is the input signal and W is the window function. Eq. (11) formulates the complex-valued *N*-point discrete STFT from which the *N*-point discrete spectrogram, $Spec(X(n,k))$, can be computed. We used *N*=256 samples long *Hanning* window with 128 samples overlap. Eqs. (12) and (13) formulate the spectral and classification loss functions, respectively.

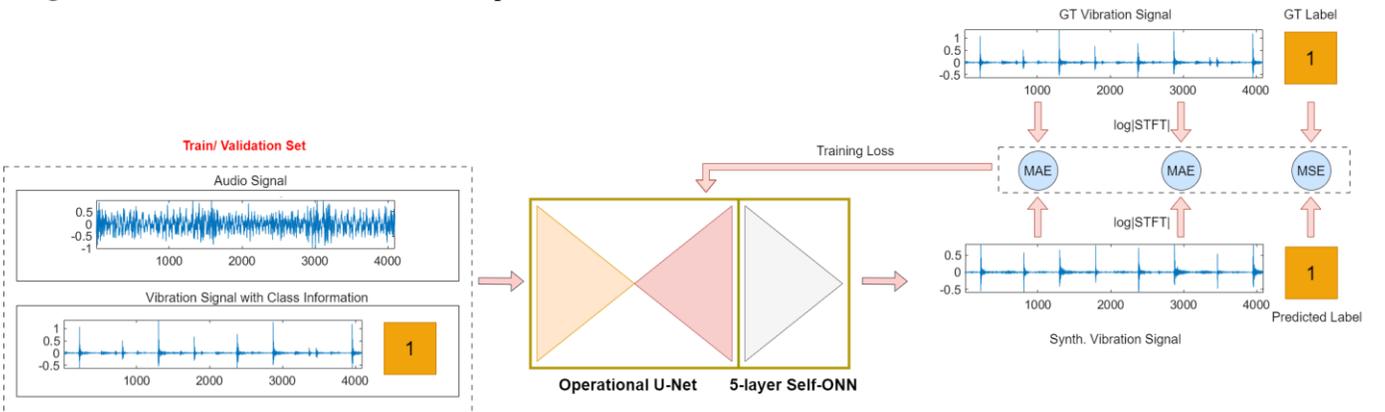

**Figure 4** The training scheme of the cascaded network model for the proposed sound-to-vibration transformation.

$$STFT[X, w, n] = X(n, w) = \sum_m X[m]W[n-m]e^{-jwm} \quad (10)$$

$$X(n,k) = X(n,w)|_{w=\frac{2\pi k}{N}} \rightarrow Spec(X(n,k)) = |X(n,k)|^2 \quad (11)$$

$$Loss_{STFT} = \|STFT(Y) - Synth(X)\|_1 \quad (12)$$

$$Loss_{class} = \frac{1}{N}\sum_{i=0}^{N}(C(Y) - C(Synth(X)))^2 \quad (13)$$

where, $C(Y)$ and $C(Synth(X))$ are the class labels of the actual and synthesized vibration signals, respectively. Finally, the overall objective function used for training combines all loss functions and can be expressed as follows:

$$Loss_{total} = Loss_{class} + \lambda(Loss_{Time} + Loss_{STFT}) \quad (14)$$

where $\lambda$, is the weight parameter that balances the temporal and spectral loss with classification loss.

### III. EXPERIMENTAL RESULTS

This section will first introduce the benchmark motor bearing fault dataset that was used in this study. Then, the experimental setup for evaluating the proposed sound-to-vibration transformer will be discussed. In Section III.C, quantitative and qualitative evaluations of the results and discussions, especially for the real use-case scenarios, are provided. Finally, in Part III.D, the computational complexity of the proposed approach is examined in-depth.

*A. Qatar University Dual-Machine Bearing Fault Benchmark Dataset: QU-DMBF*

The benchmark dataset utilized in this study was established by Qatar University researchers using 2 different electric machines (Machine A and Machine B). The experimental setup is given in Figure 1 which illustrates the orientation of the sensors and the installation of two machines. The configuration for Machine-A consists of a 3-phase AC motor, two double-row bearings, and a shaft rotating at a maximum speed of 2840 RPM. A spring mechanism placed a 6 kips radial load on the shaft and bearing. PCB accelerometers (352C33 high sensitivity Quartz ICP) mounted on the bearing housing. It weighs 180 kg and is 100x100x40cm. The working conditions for Machine-A are based on the following:

- 19 different bearing configurations: 1 healthy, 18 fault cases: 9 with a defect on the outer ring, and 9 with a defect on the inner ring. The defect sizes vary from 0.35mm to 2.35mm.
- 5 different accelerometer localization: 3 different positions and 2 different directions (radial and axial)
- 2 different load (force) levels: 0.12 kN and 0.20 kN.
- 3 different speeds: 480, 680, and 1010 RPM.

We collected data for 270 seconds for each operating circumstance for a healthy bearing, and 30 seconds for each faulty bearing case. This results in a total time of 30 x 18 x 5 x 2 x 3 = 16,200 seconds of data measurement. The sound was also simultaneously recorded with the same sampling frequency as the vibration data.

In contrast, Machine B's design consists of a DC motor, two single-row bearings, and a shaft with a constant rotating speed of 2000 RPM. A spring mechanism installed a 6 kips radial load on the shaft and bearing. PCB accelerometers (353B33 high sensitivity Quartz ICP) mounted on the bearing housing. It weighs 3.5 kg, and the configuration measures 165x82x63 cm. The working conditions for Machine B vary as follows:

- 19 different bearing configurations: 1 healthy, 9 with a defect on the outer ring, and 9 with a defect on the inner ring. The defect sizes vary from 0.35mm to 2.35mm.
- 6 different accelerometer positions.
- A fixed load (force) of 0.40 kN.
- 5 different speeds: 240, 360, 480, 700, and 1020 RPM.

270 seconds of vibration/sound data for each operating condition for a healthy bearing are available in this dataset. As a result, the total time of the healthy bearing vibration data is 270 x 6 x 1 x 5 = 8,100 seconds. 30 seconds of vibration/sound data for each working condition for each faulty bearing are available. This results in a 2:1 ratio of the faulty to healthy data, with a total time of 30 x 18 x 6 x 1 x 5 = 16,200 seconds. As a result, the dataset for machine B lasts 24,300 seconds in total (6.75 hours). The sound of each machine was simultaneously recorded with the same sampling frequency as the vibration data.

As opposed to the challenges of vibration data collection (i.e., see Figure 1), there is a crucial advantage for the sound signal acquisition as such a location sensitivity does not exist. This has been confirmed in a recent study[25] where even a DL classifier trained on the data acquired by one sensor may fail to detect certain faults in another's data. The same study has further shown that the most reliable vibration data for fault detection is acquired from the closest accelerometer to the bearings, i.e., accelerometer-1 for both machines. So, we have selected this accelerometer for training the transformers of both machines and used them to synthesize the corresponding vibration signal, which is then evaluated with the actual vibration signal. The QU-DMBF is now publicly shared in [42] and [43] to serve as the dual-machine bearing fault detection benchmark for the research community.

*B. Experimental Setup*

For the training of the proposed network model, the 1D Operational U-Net and the Self-ONN classifier, a batch size of 8, and a maximum of 1000 Back-Propagation (BP) iterations are used for all experiments. The Adam optimizer with an initial learning rate of $10^{-4}$ is used via BP. The parameter $\lambda$ in Eq (14) is set to 100. The first 2100 seconds of sound signals and their vibration counterparts are used for training, and the next 800 seconds of data are used for the validation set. For both machines, the data partition with one of the speed settings is spared for testing.

The fault detector is a compact 1D Self-ONN model with 5 operational layers and 2 dense layers. It has 32 neurons in the hidden dense layer and 16 neurons in each of the hidden operational layers. For this binary classification task, the output layer has two neurons. The input layer neuron records a 1-second vibration segment. In all layers, the "tanh" nonlinear activation function is employed. The operational layers' kernel sizes are set at 81, 41, 21, 7, and 7 respectively. Operational layers have corresponding strides of 8, 4, 2, 2, and 2. The ADAM optimizer is used with the initial learning factor, ε, set to $10^{-4}$ for the Back-Propagation (BP) training. The loss function is the Mean-Squared-Error (MSE), and the maximum number of BP iterations (epochs) is 50. We implemented both transformer and fault detector networks using the FastONN library [29] based on PyTorch. The benchmark dataset, our results, and the optimized PyTorch implementation of the proposed approach are now publicly shared with the research community [42].

## C. Results

This section presents quantitative and qualitative (visual) evaluations of the proposed methodology. To compute the quantitative results, several experiments were conducted over both real and synthesized vibration data using the Self-ONN classifier. The results of the experiments were evaluated using several common performance metrics, including accuracy, sensitivity (recall), positive predictivity (precision), and the F1-Score. Four different training and testing scenarios were selected to validate the effectiveness of the proposed approach. In each scenario, the classifier was trained and tested with non-overlapping real and synthesized vibration signals for two independent machines. The quantitative results are presented in Table 1. To provide a clear representation of these scenarios, abbreviations were used in the table for the *synthesized* and *real* data for Machine-A and Machine-B as SA, RA, SB, and RB, respectively where RA and RB correspond to *real* vibration signals. The 2$^{nd}$ and 4$^{th}$ row of the given table shows the classification performance of the proposed model over the *synthesized* vibration signals, SA and SB, respectively.

**Table 1** The fault detection performance over real (RX) and synthesized (SX) vibration data for machine X.

| Train Data | Test Data | Accuracy (%) | Healthy | | | Faulty | | |
|---|---|---|---|---|---|---|---|---|
| | | | Sensitivity (%) | Precision (%) | F1-Score (%) | Sensitivity (%) | Precision (%) | F1-Score (%) |
| RA | RA | 99.70 | 100 | 99.12 | 99.56 | 99.54 | 100 | 99.77 |
| **RA** | **SA** | 99.76 | 100 | 99.12 | 99.56 | 99.51 | 100 | 99.76 |
| RB | RB | 97.56 | 93.67 | 99.55 | 96.52 | 99.76 | 96.54 | 98.12 |
| **RB** | **SB** | 97.17 | 98.51 | 93.43 | 95.90 | 96.48 | 99.22 | 97.83 |

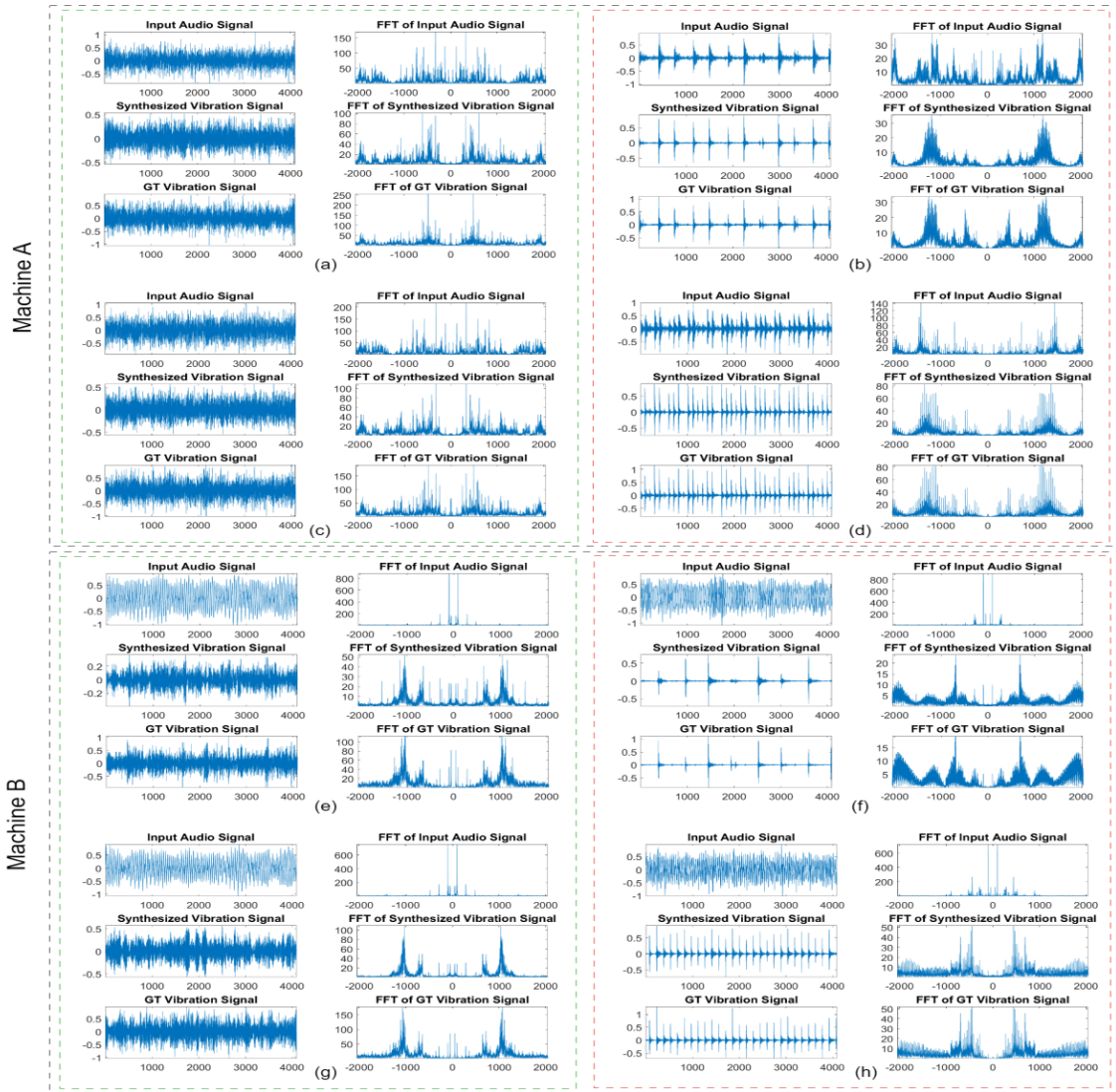

**Figure 5** 8 sets of sound-to-vibration transformation results from both machines.

Overall, fault detection results presented in Table 1 indicate that the models trained over the real vibration data, and tested over both real and synthesized data perform equally well with a high-detection performance. When we examine the individual test results, one can observe that results for Machine A are almost identical for Machine B, and the test results over real and synthesized data may differ maximum of only 0.4%, which is negligible. This is strong evidence that the proposed model can transform sound to such vibration signals that are very similar to their real counterparts and thus, the pre-trained detector can achieve a fault detection performance almost identical to the one with the real vibration data.

For the qualitative evaluation, Figure 5 presents 8 sets of sound, real and synthesized signals in both time and frequency domains corresponding to both healthy and faulty data from both machines. Once again, the results show that the synthesized vibration signals are quite similar to their real counterparts regardless of the data class (healthy or faulty). In particular, a close inspection of the spectral representation shows that the proposed transformation can indeed synthesize such signals that share the same spectral signatures of their real counterparts, i.e., the spectral peak locations representing the major spectral components in both healthy and faulty vibration data perfectly match. It is worth mentioning that the two machines produce entirely different levels of sound as a result of their varied sizes; however, this seems no effect on the transformation outputs. When we take a closer look at the signals of Machine A, one can discover that the input audio signal patterns have similarities to those of the vibration signals. The proposed approach, however, suppresses the high-frequency peaks and amplifies the low-frequency peaks in accordance with the actual spectrum.

### D. Computational Complexity Analysis

The network size, total number of parameters (PARs), and inference time for each network configuration are computed in this section. Detailed formulations of the PARs calculations for Self-ONNs are available in [29]. A 2.2 GHz Intel Core i7 computer with 16 GB of RAM and an NVIDIA GeForce RTX 3080 graphics card was used for all experiments. The FastONN library [29] and Pytorch are used to implement the 1D Op-UNet network. The proposed network has a total of 377K parameters. The processing time to generate a 1-second vibration segment takes around 6.5 msec for a single CPU implementation. This shows that the proposed transformation can achieve 150 times faster than the real-time requirements with a single CPU, indicating the potential of a real-time implementation even on low-cost, low-power hardware.

## IV. CONCLUSIONS

Efficient detection of bearing faults is crucial for predictive maintenance in various industries. In this study, we propose a novel sound-to-vibration transformation method that can synthesize realistic vibration signals regardless of the working conditions or fault status. It has been shown in this paper that a simple audio recorder (e.g., a mobile phone) and the proposed machine learning model are sufficient for an accurate fault detection. The quantitative results show that the fault detection accuracy difference achieved using synthesized and real data is less than 0.5%, which is negligible. Regardless of the motor type (AC/DC), size, fault type/severity, and sound level, the results demonstrate that the proposed method can transform the sound signal to synthesize the corresponding (real) vibration signal. Therefore, the proposed approach makes motor health monitoring significantly more practical and accessible, as it eliminates the need for any vibration sensor. It is also highly efficient, inexpensive, and robust because all of the aforementioned challenges and drawbacks associated with using accelerometers for data acquisition can effectively be eliminated with a simple sound recorder. As a result, the proposed approach has the potential to revolutionize the field of predictive maintenance and make the process more accessible and practical for various other applications, e.g., mechanical fault detection on vehicles or any moving platform in general.

Our future study will focus on performing fault recognition to further identify the type of defect, estimate its severity, and pinpoint the defect's location for comprehensive real-time health monitoring in a *sensorless* fashion. Finally, zero-shot fault detection [25] using only the sound signal will be another crucial objective for our future research.

## REFERENCES

[1] Z. Gao, C. Cecati and S. X. Ding, "A survey of fault diagnosis and fault-tolerant techniques—part I: fault diagnosis with model-based and signal-based approaches," *IEEE Trans. Ind. Electron.*, vol. 62, no. 6, pp. 3757–3767, 2015.

[2] X. Dai and Z. Gao, "From model, signal to knowledge: a data-driven perspective of fault detection and diagnosis," *IEEE Trans. Ind. Informat.*, vol. 9, no. 4, pp. 2226–2238, 2013.

[3] F. Filippetti, A. Bellini, and G. A. Capolino, "Condition monitoring and diagnosis of rotor faults in induction machines: state of art and future perspectives," in Proc. IEEE WEMDCD, Paris, Mar. 2013, pp. 196–209.

[4] S. Sassi, B. Badri and M. Thomas, 2007. A Numerical Model to Predict Damaged Bearing Vibrations Journal of Vibration and Control, Vol. 13, No. 11, pp. 1603-1628.

[5] R.R. Schoen, T.G. Habetler, F. Kamran, and R. G. Bartheld, "Motor bearing damage detection using stator current monitoring," *IEEE Trans. Ind. Appl.*, vol. 31, pp. 1274-1279, Dec. 1995.

[6] M.H. Benbouzid, "A Review of Induction Motors Signature Analysis as a Medium for Faults Detection" in IEEE Transactions on Industrial Eletronics, 2000, Vol. 47, NO. 5.

[7] J. Pons-Llinares, J.A. Antonino-Daviu, M. Riera-Guasp, S. B. Lee, and T.J. Kang and C. Yang "Advanced induction motor rotor fault diagnosis via continuous and discrete time–frequency tools," IEEE Trans. Ind. Electron., vol. 62, no 3, pp. 1791 – 1802, Mar. 2015.

[8] L. Eren, M. J. Devaney, "Bearing damage detection via wavelet packet decomposition of the stator current", IEEE Trans. Instrum. Meas., vol. 53, no. 2, pp. 431–436, 2004.

[9] S. Nandi, H. A. Toliyat, and X. Li, "Condition monitoring and fault diagnosis of electrical motors: A review," IEEE Trans. Energy Conv., vol. 20, no. 4, pp. 719–729, Dec. 2005.

[10] V.T. Do, and U.-P. Chong, (2011). Signal model-based fault detection and diagnosis for induction motors using features of vibration signal in two-dimension domain. *Strojniški vestnik*, *57*(9), 655-666. https://doi.org/10.5545/sv-jme.2010.162

[11] A. Bellini, F. Filippetti, C. Tassoni and G. A. Capolino, "Advances in diagnostic techniques for induction machines," *IEEE Trans. Ind. Electron.*, vol. 55, no. 12, pp. 4109–4125, 2008.

[12] X.S. Lou, K.A. Loparo, "Bearing fault diagnosis based on wavelet transform and fuzzy inference," Mech. Syst. Signal Process., vol. 18, pp. 1077-1095, 2004.

[13] W. Sun, S. Shao, R. Zhao, R. Yan, X. Zhang, and X. Chen, "A sparse auto-encoder-based deep neural network approach for induction motor faults classification," Measurement, vol. 89, pp. 171–178, 2016.

[14] M. Xia, T. Li, L. Liu, L. Xu, and C. de Silva, "An intelligent fault diagnosis approach with unsupervised feature learning by stacked denoising autoencoder," IET Sci. Meas. Technol., vol. 11, pp. 687-695, 2017.

[15] F. Jia, Y. Lei, L. Guo, J. Lin, S. Xing, A neural network constructed by deep learning technique and its application to intelligent fault diagnosis of machines, Neurocomputing, vol. 272, pp. 619–628, 2018.

[16] Z. Chen, C. Li, and R.-V. Sanchez, "Gearbox fault identification and classification with convolutional neural networks," Shock Vib., vol. 2015, pp. 1-10, 2015.

[17] O. Janssens et al., "Convolutional neural network based fault detection for rotating machinery," J. Sound Vib., vol. 377, pp. 331–345, 2016.

[18] X. Guo, L. Chen, and C. Shen, "Hierarchical adaptive deep convolution neural network and its application to bearing fault diagnosis," Measurement, vol. 93, pp. 490–502, 2016.

[19] S. Shao, P. Wang, R. Yan, "Generative adversarial networks for data augmentation in machine fault diagnosis," Computers in Industry, vol. 106, pp. 85-93, 2019.

[20] S. Liu,; J. Xie, C. Shen,X. Shang, D. Wang, Z. Zhu, "Bearing Fault Diagnosis Based on Improved Convolutional Deep Belief Network," Appl. Sci. 2020, 10, 6359.

[21] S. Zhang, S. Zhang, B. Wang and T. G. Habetler, "Deep Learning Algorithms for Bearing Fault Diagnostics—A Comprehensive Review," in IEEE Access, vol. 8, pp. 29857-29881, 2020, doi: 10.1109/ACCESS.2020.2972859.

[22] O. Avci, O. Abdeljaber, S. Kiranyaz, S Sassi, A. Ibrahim and M. Gabbouj, 2022. One-dimensional convolutional neural networks for real-time damage detection of rotating machinery. In Rotating Machinery, Optical Methods & Scanning LDV Methods, Volume 6: Proceedings of the 39th IMAC, A Conference and Exposition on Structural Dynamics 2021 (pp. 73-83).

[23] T. Ince, S. Kiranyaz, L. Eren, M. Askar, M. Gabbouj, "Real-time motor fault detection by 1-D convolutional neural networks," IEEE Trans. Ind. Electron., vol. 63, no. 11, pp. 7067–7075, 2016.

[24] T. Ince, J. Malik, O.C. Devecioglu, S. Kiranyaz, O. Avci, L. Eren and M. Gabbouj, "Early Bearing Fault Diagnosis of Rotating Machinery by 1D Self-Organized Operational Neural Networks," in IEEE Access, vol. 9, pp. 139260-139270, 2021, doi: 10.1109/ACCESS.2021.3117603.

[25] S. Kiranyaz and O. C. Devecioglu and A. Alhams and S. Sassi and T. Ince and O. Abdeljaber and O. Avci and M. Gabbouj (2022). Zero-Shot Motor Health Monitoring by Blind Domain Transition. ArXiv, 2212.06154.

[26] F. Mohd-Yasin, C. E. Korman, D. J. Nagel, Measurement of noise characteristics of MEMS accelerometers, Solid-State Electronics,Volume 47, Issue 2,2003,Pages 357-360,ISSN 0038-1101

[27] S. Kiranyaz, J. Malik, H. B. Abdallah, T. Ince, A. Iosifidis, M. Gabbouj, "Self-Organized Operational Neural Networks with Generative Neurons,", Neural Networks, Volume 140, 2021, Pages 294-308, ISSN 0893-6080, https://doi.org/10.1016/j.neunet.2021.02.028.

[28] J. Malik, S. Kiranyaz, M. Gabbouj, "Self-Organized Operational Neural Networks for Severe Image Restoration Problems", Neural Networks (Elsevier), vol. 135, pp. 201-211, Jan. 2021. https://doi.org/10.1016/j.neunet.2020.12.014

[29] J. Malik, S. Kiranyaz, M. Gabbouj, "FastONN--Python based open-source GPU implementation for Operational Neural Networks", arXiv preprint arXiv:2006.02267, Mar. 2020.

[30] S. Kiranyaz, T. Ince, A. Iosifidis and M. Gabbouj, "Operational Neural Networks," Neural Computing and Applications (Springer-Nature), Mar. 2020. DOI: https://doi.org/10.1007/s00521-020-04780-3

[31] S. Kiranyaz, J. Malik, H. B. Abdallah, T. Ince, A. Iosifidis, M. Gabbouj, "Exploiting Heterogeneity in Operational Neural Networks by Synaptic Plasticity", Neural Computing and Applications (Springer-Nature), pp. 1-19, Jan. 2021. https://doi.org/10.1007/s00521-020-05543-w

[32] S. Kiranyaz, T. Ince, A. Iosifidis and M. Gabbouj, "Progressive operational perceptrons," Neurocomputing, vol. 224, pp.142–154, 2017.

[33] D.T. Tran, S. Kiranyaz, M. Gabbouj and A. Iosifidis, "Progressive operational perceptron with memory," Neurocomputing 379, pp. 172-181, 2020.

[34] O. C. Devecioglu, J. Malik, T. Ince, S. Kiranyaz, E. Atalay, M. Gabbouj, "Real-Time Glaucoma Detection From Digital Fundus Images Using Self-ONNs", IEEE Access, Vol. 9, pp. 140031-140041, Sep. 2021. DOI: 10.1109/ACCESS.2021.3118102

[35] J. Malik, O. C. Devecioglu, S. Kiranyaz, T. Ince, M. Gabbouj, "Real-Time Patient-Specific ECG Classification by 1D Self-Operational Neural Networks," IEEE Trans. on Biomedical Engineering, Dec. 2021. DOI: 10.1109/TBME.2021.3135622

[36] M. Gabbouj, S. Kiranyaz, J. Malik, M. U. Zahid, T. Ince, M. Chowdhury, A. Khandakar, and A. Tahir, "Robust Peak Detection for Holter ECGs by Self-Organized Operational Neural Networks", in IEEE Trans. on Neural Networks and Learning Systems, April 2022.

[37] M. Uzair, S. Kiranyaz and M. Gabbouj, "Global ECG Classification by Self-Operational Neural Networks with Feature Injection", IEEE Transactions on Biomedical Engineering, pp. 1-12, July 2022. DOI: 10.1109/TBME.2022.3187874.

[38] M Soltanian, J Malik, J Raitoharju, A Iosifidis, S Kiranyaz, M Gabbouj, "Speech Command Recognition in Computationally Constrained Environments with a Quadratic Self-Organized Operational Layer", in Proc. of Int. Joint Conference on Neural Networks (IJCNN), pp. 1-6, July 2021.

[39] J. Malik, S. Kiranyaz, M. Gabbouj, "BM3D vs 2-Layer ONN", in Proc. of IEEE Int. Conference on Image Processing (ICIP), Sep. 2021. DOI:10.1109/ICIP42928.2021.9506240

[40] S. Kiranyaz, O. C. Devecioglu, T. Ince, J. Malik, M. Chowdhury, T. Hamid, R. Mazhar, A. Khandakar, A. Tahir, T. Rahman, and M. Gabbouj, "Blind ECG Restoration by Operational Cycle-GANs", IEEE Transactions on Biomedical Engineering, May 2022.

[41] J. Malik, S. Kiranyaz, M. Gabbouj, "Operational vs Convolutional Neural Networks for Image Denoising", arXiv:2009.00612, Sep. 2020.

[42] Sound to Vibration Transformation for Sensorless Motor Health Monitoring, Version 1.0, Source code. [Online]. Available: https://github.com/OzerCanDevecioglu/Sound-to-Vibration-Transformation-for-Sensorless-Motor-Health-Monitoring

[43] Zero Shot Bearing Fault Detection by Blind Domain Transition, Version 1.0, Source code. [Online]. Available: https://github.com/OzerCanDevecioglu/Zero-Shot-Bearing-Fault-Detection-by-Blind-Domain-Transition.